\documentclass[12pt]{article}
\usepackage[latin9]{inputenc}
\usepackage{color}
\usepackage{array}
\usepackage{bm}
\usepackage{multirow}
\usepackage{amsmath}
\usepackage{amssymb}
\usepackage{graphicx}
\usepackage{setspace}
\usepackage{esint}
\usepackage[authoryear]{natbib}
\PassOptionsToPackage{normalem}{ulem}
\usepackage{ulem}

\makeatletter

\providecommand{\tabularnewline}{\\}

\usepackage{psfrag}\usepackage{epsf}\usepackage{enumerate}\usepackage{url}
\usepackage{soul}

\usepackage{color}\usepackage{bm}

\newcommand{\ignore}[1]{}

\makeatother
\date{}

\begin{document}

\title{\textbf{Spatial process decomposition for quantitative imaging biomarkers
using multiple images of varying shapes}}

\singlespacing
\author{\normalsize{ShengLi Tzeng$^{1}$, Jun Zhu$^{2,3}$\footnote{Corresponding author. Email address: jzhu@stat.wisc.edu},  
Amy Weisman$^{4}$, 
Tyler Bradshaw$^{4}$ and Robert Jeraj$^{4,5}$ } \\ 
$^{1}$\small{Department of Applied Mathematics, National Sun Yat-sen University, Taiwan}\\ 
$^{2}$\small{Department of Statistics,  University of Wisconsin-Madison, USA} \\ 
$^{3}$\small{Department of Entomology,  University of Wisconsin-Madison, USA} \\ 
$^{4}$\small{Department of Medical Physics, University of Wisconsin Madison, USA} \\
$^{5}$\small{Department of Human Oncology, University of Wisconsin Madison, USA}  }

\maketitle

\doublespacing
\begin{abstract}
Quantitative imaging biomarkers (QIB) are extracted from medical images in radiomics for a variety of purposes including noninvasive disease detection, cancer monitoring, and precision medicine. The existing methods for QIB extraction tend to be \textit{ad-hoc} and not reproducible. In this paper, a general and flexible statistical approach is proposed for handling up to three-dimensional medical images in an objective and principled way. In particular, a model-based spatial process decomposition is developed where the random weights are unique to individual patients for component functions common across patients. Model fitting and selection are based on maximum likelihood, while feature extractions are via optimal prediction of the underlying true image. A simulation study evaluates the properties of the proposed methodology and for illustration, a cancer image data set is analyzed and QIBs are extracted in association with a clinical endpoint. 
\end{abstract}

\vfill{}

%
%
\section{Introduction}\label{sec:intro}

With the advance of imaging techniques, radiomics has emerged as a field in medicine that concerns extraction of quantitative imaging biomarkers (QIBs) from medical images, finding a variety of applications such as noninvasive disease diagnosis, cancer monitoring, treatment response assessment, and precision medicine \citep[see, e.g.,][]{gillies2015radiomics}.
While QIBs overcome the limitations of traditional tools including subjective visual interpretation with inter-observer variability that often occur in medical image assessment, the existing methods in radiomics could benefit more from statistical approaches in connection to the subsequent data analysis involving QIBs. 
In this paper, we propose a novel statistical method for extracting potential QIBs, providing a general framework for dealing with up to three-dimensional (3D) medical images in an objective and statistically more rigorous way.

The current procedures for developing QIBs are rather complicated, involving image acquisition, region of interest (ROI) identification and segmentation, potential biomarker quantification and extraction, and association with clinical endpoints and validation.
For example, \citet{fares2011individuals} followed a QIB procedure that started with acquisition by digitizing dental X-ray panoramic images into 2621$\times$1244 pixels, and then used contrast enhancement and morphological operations for segmenting jaws and teeth.
The contours of ROIs were delineated based on the segmentations and three different features were extracted (i.e., dental work performed on the individual, number of dental screws, and detection of missing teeth), followed by testing the performance of the extracted features. 

Texture analysis (TA) is a commonly used technique for potential
biomarker quantification and extraction, which systematically constructs high-throughput information from ROIs \citep[see, e.g.,][]{xie2008galaxy, parekh2016radiomics}. 
However, this technique has multiple challenges.
First, the number of extracted TA features is often much larger than the number of patients in a study, in which case popular classical statistical tools (e.g., linear regression, Cox proportional hazard regression) can not be directly applied. 
Appropriate dimension reduction methods are therefore important for not only mitigating over-fitting but also alleviating high correlations among a large number of high-throughput features. 
In practice, however, dimension reduction is not always performed. For example, \citet{parmar2015machine}
reviewed several univariate filtering methods which associate a clinical endpoint of interest with one feature at a time, not considering the dependency between features. 
If several features have high correlations with one important feature, these filtering methods are likely to
select most of these features and the resulting strong collinearity would make many statistical tools challenging to use. 
Thus, despite their popularity, filtering methods are not satisfactory in such context. 
A simple alternative to filtering methods is to consider commonly-used features in the literature, leading
to more interpretable analysis. 
Unfortunately, commonly-used features are not necessarily the more useful ones and worse yet, focusing
on well-known features impedes the discovery of other viable biomarkers. 

Second, the existing TA techniques were developed largely for rectangular images but have been widely adopted in cancer studies where the tumor ROIs are non-rectangular. 
\citet{zand2015texture} identified such a shape issue arising from non-rectangular two-dimensional (2D) images and reviewed \textit{ad hoc} remedies such as embedding a non-rectangular ROI in a rectangular image with blank pixels surrounding the ROI; however, this approach becomes even more problematic for 3D cases with a much larger portion of blank voxels in the cube where the ROI is embedded. 
In addition, the images to be analyzed are not conformed to the same shape or size across patients, as individual patients' ROIs are unique. 
A naive way would be to use the pixel-wise (or voxel-wise) average of TA features in a local rectangle (or cube), but would potentially lose much information in the spatial structures. 
How to extract important spatial characteristics from ROIs with different shapes and sizes is a formidable problem that seems largely unaddressed. 

A third challenge is that many TA techniques require binning the intensity values observed on a medical image into distinct categories, known as quantization, for constructing local-pattern matrices (e.g., neighboring gray-level dependence and co-occurrence) in the extraction of QIBs. 
\citet{hatt2017characterization} demonstrated by examples that quantization has a large impact on many TA features. 
Yet there is little consensus on the number of bins and the binning method to use, resulting in arbitrary quantization and hindering the reproducibility of studies. 

In light of the aforementioned challenges with TA, we propose a new statistical method applicable for extracting nearly uncorrelated features for the intensities in ROIs. 
A key strategy is model-based spatial process decomposition, which formulates the image data as linear combinations of common spatial processes, building upon previous work for similar objectives. 
For example, the wavelets and principal component analysis \citet{gupta2006wavelet} developed for hyperspectral images produced multi-resolution features capturing important spatial patterns, but unfortunately was not easily applicable to non-rectangular ROIs. 
\citet{wang2017regularized} proposed a regularized principal component analysis method for irregularly shaped 2D data, although each patient was required to share the same sampling locations. 
\citet{zhou2014principal} used functional mixed-effects models for 2D functions by employing bivariate splines on triangulations, which handles varying shapes. 
However, the triangulation requires that each triangulated subregion contains at least a few data locations, which makes its extension to 3D images a challenge.

The novelty of our proposed methodology compared with the previous work is its ability to deal with 1D time point, 2D pixel, and 3D voxel data within a same general framework of spatial random-effects models.
Although the spatial random-effects model is a popular geostatistical tool, the purpose in geostatistics is largely spatial regression or kriging at unsampled locations \citep[see, e.g.,][]{wikle2010[clm], TzengHuang2017}, whereas our main goal is to find the important building blocks for the observed image intensities. 
Further, for the most part, geostatistical data are 1D or 2D and there are no independent replicates, while in radiomics, the images are 2D or 3D from multiple patients.
Our innovation here is to let the random effects in our model carry weights, unique to individual patients, for the common building blocks, and an orthogonalization of these random effects produces nearly uncorrelated biomarkers. 
Finally, our proposed methodology has the advantage that the medical images require no quantization and the ROIs could have different shapes.

The remainder of this paper proceeds as follows. In Section~\ref{sec:model}, we propose a data model and a latent model. The methodology for parameter estimation, model selection, and feature extraction is given in Section~\ref{sec:method}. Section~\ref{sec:simu} presents a simulation study evaluating the properties of
our proposed method. For illustration, a data example is presented in Section~\ref{sec:example}, followed by conclusions and a discussion in Section~\ref{sec:conclusion}.

%
%
\section{Model}\label{sec:model}

\subsection{Data Model}\label{sec:modeld}

\begin{figure}[t]
\begin{centering}
\begin{tabular}{ccc}
\includegraphics[scale=0.5]{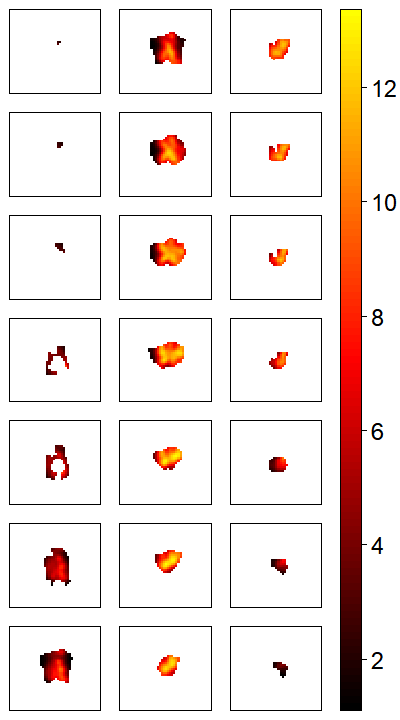} & & \includegraphics[scale=0.5]{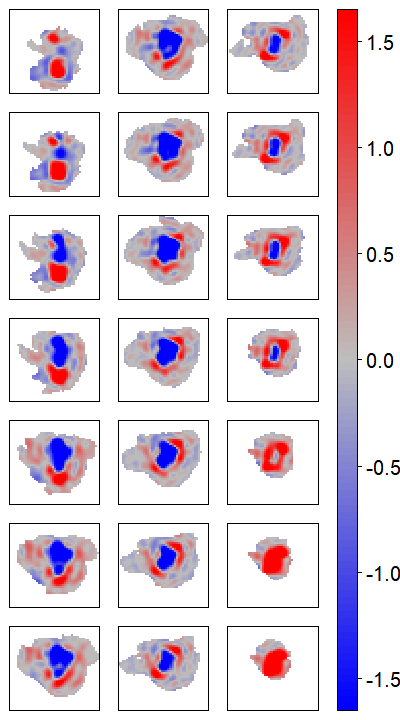}\tabularnewline
(a) & & (b)\tabularnewline
\end{tabular}
\par
\end{centering}
\caption{An example of 21 consecutive transverse slices of fluorodeoxyglucose on a positron emission tomography (FDG PET) image in a 3D region of interest for a canine patient (left panel) and one of the estimated component functions $\hat{g}_{h}(\cdot)$ (right panel).}
\label{fig:zandg}
\end{figure}

Let $N\geq1$ denote the number of patients and each patient has a 2D (or 3D) image of ROI with observed intensities across pixels (or voxels) for an index under study (e.g., fluorodeoxyglucose level in Figure~\ref{fig:zandg}). 
Let $n_j$ denote the number of observation locations, $\bm{s}_{j,1},\ldots,\bm{s}_{j,n_{j}}$, generally the centroids of pixels or voxels, for the $j$th patient with $j=1,\ldots,N$.
Let the union of the observation locations among all the patients be denoted as $D=\{\bm{s}_{1},\dots,\bm{s}_{n}\}=\cup_{j=1}^{N}\{\bm{s}_{j,1},\ldots,\bm{s}_{j,n_{j}}\}$, where $\bm{s}_{1},\dots,\bm{s}_{n}$ are $n$ distinct observation locations and $D\subset\mathbb{R}^{d}$ is the collection of locations in a $d$-dimensional common spatial domain shared across the $N$ patients for $d=2$ or $3$. 
Thus, each patient has $n$ potential observation locations, of which $n_{j}\leq n$ locations have observations and the other $n-n_{j}$ locations have no observations for the $j$th patient. 
When $N=1$, there is just one patient and $n_{j}=n$; that is, we only have a single realization, which is common in geostatistics but not in radiomics. 

Next, for the $j$th patient where $j=1,\dots,N$, we define an $n_{j}\times n$ incidence matrix $\mathbf{O}_{j}$, whose $(k,i)$th element is 1 if $\bm{s}_{j,k}=\bm{s}_{i}$ and is 0 otherwise. 
Let $z_{j}(\bm{s}_{i})$ denote the potentially observable image intensity at location $\bm{s}_{i}$ for $i=1,\ldots,n$ and let $\bm{z}_{j}=\left(z_{j}(\bm{s}_{1}),\dots,z_{j}(\bm{s}_{n})\right)'$ denote the vector of image intensities at all the locations with and without observations. 
Then, $\mathbf{O}_{j}\bm{z}_{j}$ is the vector of observed image intensity data from the $j$th patient. 
In the context of QIB, we define a latent variable $y_{j}(\bm{s})$ to represent the \textit{true} image intensity at location $\bm{s}\in D$ and let $\bm{y}_{j}=(y_{j}(\bm{s}_{1}),\dots,y_{j}(\bm{s}_{n}))'$ denote the vector of true image intensities at all the locations for the $j$th patient.

In what follows, we assume that the image intensity data for the $j$th patient are observed according to $z_{j}(\bm{s}_i)=y_{j}(\bm{s}_i)+\varepsilon_{j}(\bm{s}_i)$, where $\varepsilon_{j}(\bm{s}_i)$ is an error corresponding to white noise, for $i=1,\ldots,n$ and $j=1,\ldots,N$. 
Let $\bm{\varepsilon}_{j}=\left(\varepsilon_{j}(\bm{s}_{1}),\ldots,\varepsilon_{j}(\bm{s}_{n})\right)^{\prime}$ denote a vector of errors. 
For $j=1,\ldots,N$, we have 
\begin{equation}
\mathbf{O}_{j}\bm{z}_{j}=\mathbf{O}_{j}(\bm{y}_{j}+\bm{\varepsilon}_{j}),
\label{eq:uni-obs-equation-z}
\end{equation}
where $\bm{\varepsilon}_{j}\sim\mathcal{N}(\bm{0},\sigma^{2}\bm{I}_{n})$ is assumed to be uncorrelated with $\bm{y}_{j}$, and $\bm{\varepsilon}_{j}$ is assumed to be uncorrelated with $\bm{\varepsilon}_{j^{*}}$ for $j\neq j^{*}$. 

\subsection{Latent Model}\label{sec:modell}

For $j=1,\ldots,N$, we model the $j$th latent variable $y_{j}(\bm{s})$ by
\begin{equation}
y_{j}(\bm{s})=\mu_{j}(\bm{s})+\eta_{j}(\bm{s}), 
\label{eq:uni-latent-y}
\end{equation}
where  $\mu_{j}(\bm{s})$ is a deterministic mean function and $\eta_{j}(\cdot)$ is a mean-zero spatial process with a spatial covariance function $C(\bm{s},\bm{s}^{*})=\mathrm{cov}(y_{j}(\bm{s}),\, y_{j}(\bm{s}^{*}))$ for $\bm{s},\bm{s}^{*}\in D$.  
By a truncated version of the Karhunen-Lo{\'e}ve expansion \citep[see, e.g.,][]{alexanderian2015brief}, 
we model $\eta_{j}(\bm{s})$ as a linear combination of $H$ component functions $g_h(\bm{s})$ such that
\begin{equation}
\eta_{j}(\bm{s})=\sum_{h=1}^{H}\theta_{jh}g_{h}(\bm{s}),
\label{eq:uni-latent-eta}
\end{equation}
where orthogonality holds for $g_h(\cdot)$ such that $\sum_{i=1}^n g_{h}(\bm{s}_i)g_{h^{*}}(\bm{s}_i)=0$ for $h\neq h^{*}$.
Further, let $\bm{\theta}_{j}=(\theta_{j1},\dots,\theta_{jH})'$ denote a vector of random weights such that
$\bm{\theta}_{j}\sim\mathcal{N}(\bm{0},\bm{\Lambda})$, where $\bm{\Lambda}=\mathrm{diag}\{\lambda_{1},\dots,\lambda_{H}\}$ is an $H\times H$ diagonal matrix with variance components $\lambda_h$ satisfying $\lambda_{1}\geq\cdots\geq\lambda_{H}>0$, and $\bm{\theta}_{j}$ and $\bm{\theta}_{j^*}$ are uncorrelated for $j\neq j^*$. 
By \eqref{eq:uni-latent-y}--\eqref{eq:uni-latent-eta}, we have, for $\bm{s}\in D$,
\begin{equation}
y_{j}(\bm{s})=\mu_{j}(\bm{s})+\sum_{h=1}^{H}\theta_{jh}g_{h}(\bm{s}).
\label{eq:sum-of-eigen-functions}
\end{equation}
The covariance function, for $\bm{s},\bm{s}^{*}\in D$, is
\begin{equation}
C(\bm{s},\bm{s}^{*})=\textrm{cov}(y_{j}(\bm{s}),y_{j}(\bm{s}^*))=\sum_{h=1}^{H}\lambda_{h}g_{h}(\bm{s})g_{h}(\bm{s}^{*}).
\label{eq:cov-func-C}
\end{equation}

For flexible forms of the component functions, we represent $g_{h}(\bm{s})$ by a linear combination of $K$ pre-specified basis functions, denoted as $f_{1}(\bm{s}),\dots,f_{K}(\bm{s})$, for $\bm{s}\in D$. That is,
\begin{align}
g_{h}(\bm{s}) & =\sum_{k=1}^{K}u_{hk}f_{k}(\bm{s})=\boldsymbol{u}_{h}^{\prime}\bm{f}_{K}(\bm{s}),
\label{eq:sum-of-basis-functions}
\end{align}
where $\bm{f}_{K}(\bm{s})=(f_{1}(\bm{s}),\dots,f_{K}(\bm{s}))'$ denotes a vector of $K$ basis functions at location $\bm{s}$ and $\boldsymbol{u}_{h}=(u_{h1},\ldots,u_{hK})^{\prime}$ denotes a vector of coefficients for the $h$th component function. 
With $H\leq K\leq n$, the representation of the component function $g_{h}(\bm{s})$ with basis functions bears similarity to regression splines that approximate complex covariance functions by a set of piecewise polynomial bases \citep[see, e.g.,][]{eilers1996flexible, chang2010semiparam, wikle2010[clm]}.
In particular, let  $\bm{U}=[\boldsymbol{u}_{1},\ldots,\boldsymbol{u}_{H}]$ denote a $K\times H$ matrix of coefficients and let $\bm{w}_{j}=(w_{j1},\dots,w_{jK})'=\bm{U}\bm{\theta}_{j}$ denote a vector of random effects.
Since $\bm{\theta}_{j}\sim\mathcal{N}(\bm{0},\bm{\Lambda})$ in \eqref{eq:uni-latent-eta}, we have $\bm{w}_{j}\sim\mathcal{N}(\bm{0},\bm{M})$ where $\bm{M}=\bm{U\bm{\Lambda}U}^{\prime}$.
It is straightforward to show that the variance matrix $\bm{M}=\bm{U\bm{\Lambda}U}^{\prime}$ is nonnegative-definite and $\bm{w}_{j}$'s are mutually uncorrelated and also uncorrelated with $\bm{\varepsilon}_{j}$'s
in (\ref{eq:uni-obs-equation-z}). 

By (\ref{eq:sum-of-eigen-functions}) and (\ref{eq:sum-of-basis-functions}), we can rewrite $y_{j}(\cdot)$, for $j=1,\dots,N$, as
\begin{align}
y_{j}(\bm{s})= & ~\mu_{j}(\bm{s})+\sum_{k=1}^{K}w_{jk}f_{k}(\bm{s})=~\mu_{j}(\bm{s})+\bm{w}_{j}'\bm{f}_{K}(\bm{s}).
\label{eq:uni-general-form-y}
\end{align}
By \eqref{eq:uni-general-form-y}, the covariance function \eqref{eq:cov-func-C}, for $\bm{s},\bm{s}^{*}\in D$, now becomes 
\begin{equation}
C(\bm{s},\bm{s}^{*})=\textrm{cov}(y_{j}(\bm{s}),y_{j}(\bm{s}^*))=\bm{f}_{K}(\bm{s})'\bm{M}\bm{f}_{K}(\bm{s}^{*}).
\label{eq:covariance}
\end{equation}
The covariance function given in \eqref{eq:covariance} is quite flexible, as there is no need to assume stationarity and no specific parametric structure is assumed of $\bm{M}$. 

Thus, the data model \eqref{eq:uni-obs-equation-z} can be rewritten as
\begin{equation}
z_j(\bm{s}_i)=~\mu_{j}(\bm{s}_i)+\sum_{k=1}^{K}w_{jk}f_{k}(\bm{s}_i)+\varepsilon_j(\bm{s}_i)=~\mu_{j}(\bm{s}_i)+\bm{w}_{j}'\bm{f}_{K}(\bm{s}_i)+\varepsilon_j(\bm{s}_i),
\label{eq:data-model}
\end{equation}
such that the image intensity for a given pixel or voxel $\bm{s}_i\in D$ within a patient's ROI has three additive elements: (1) a patient-specific mean function $\mu_{j}(\bm{s}_i)$; (2) a linear combination of basis functions $\bm{w}_{j}'\bm{f}_{K}(\bm{s}_i)$ that captures spatial correlations between pixels or voxels with random effects and a class of pre-specified basis functions such that $\bm{w}_{j}\sim\mathcal{N}(\bm{0},\bm{M})$ where $\bm{M}=\bm{U\bm{\Lambda}U}^{\prime}$, and (3) a white noise $\varepsilon_j(\bm{s}_i)$ that captures measurement error and local heterogeneity such that $\bm{\varepsilon}_{j}\sim\mathcal{N}(\bm{0},\sigma^{2}\bm{I}_{n})$.
Since $g_{h}(\bm{s})$ and $\bm{\Lambda}$ are common across patients, $\bm{U}$, $\bm{f}_{K}(\bm{s})$, and $\bm{M}$ are also the same for all patients. 
In other words, the image data across patients are modeled as independent realizations from \eqref{eq:data-model} including those components shared among patients.

%
%
\section{Methodology}\label{sec:method}
\subsection{Multi-resolution Basis Functions}\label{sec:basis function}

With the observation locations $\{\bm{s}_{1},\ldots,\bm{s}_{n}\}$ as the control points, we construct basis functions based on thin-plate splines (TPS) kernels denoted as, 
$\psi_{i}(\bm{s})=(1/12)\|\bm{s}-\bm{s}_{i}\|^{3}$ for $d=1$, 
$\psi_{i}(\bm{s})=1/(8\pi)\|\bm{s}-\bm{s}_{i}\|^{2}\log\left(\|\bm{s}-\bm{s}_{i}\|\right)$ for $d=2$, and
$\psi_{i}(\bm{s})=-(1/8)\|\bm{s}-\bm{s}_{i}\|$ for $d=3$, where $i=1,\ldots,n$ and $\|\cdot\|$ denotes the $L^2$ norm \citep[see, e.g.,][]{ wahba1980some[TPS], green1993nonparametric}.
Let $\bm{\Psi}=\left(\psi_{i}(\bm{s}_{i^{*}})\right)_{n\times n}$ denote an $n\times n$ matrix of TPS kernels comprising $\psi_{i}(\bm{s}_{i^*})$ for $i,i^*=1,\ldots,n$.
We also define an $n\times(d+1)$ design matrix to be $\bm{X}=\left((1,\bm{s}_{1}^{\prime})^{\prime},\ldots,(1,\bm{s}_{n}^{\prime})^{\prime}\right)^{\prime}$ where the $i$th row is  $(1,\bm{s}_{i}^{\prime})^{\prime}$ for $i=1,\ldots,n$ and we let $\bm{\Omega}=\bm{I}_n-\bm{X}(\bm{X}'\bm{X})^{-1}\bm{X}'$ be an $n\times n$ projection matrix.

With the notation above, we define the basis functions $f_{k}(\bm{s})$ as follows.
Let $f_{1}(\bm{s})=1$ and $f_{k}(\bm{s})=s_{k-1}$ for $k=2,\dots,d+1$, where $\bm{s}=(s_{1},\ldots,s_{d})^{\prime}$.
For $k=d+2,\dots,n$, we form $f_{k}(\bm{s})$ by linear combinations of the TPS kernels where the coefficients are obtained from an eigen-decomposition, which essentially extracts the orthogonal subspaces in TPS but not in the space of linear functions. 
Specifically, we pre- and post-multiply $\bm{\Psi}$ by the projection matrix $\bm{\Omega}$ and then solve for the eigen-vectors $\bm{V}$ and scalars $\alpha_{k}$ that satisfy, $\bm{V}\mathrm{diag}\{\alpha_{1},\dots,\alpha_{n}\}\bm{V}'=\bm{\Omega}\bm{\Psi}\bm{\Omega}$, with $\alpha_{1}\geq\cdots\geq\alpha_{n}$.
Now, let $f_{k}(\bm{s})=\alpha_{k-d-1}^{-1}\big\{\bm{\psi}(\bm{s})-\bm{\Psi}\bm{X}(\bm{X}'\bm{X})^{-1}\bm{x}\big\}'\bm{v}_{k-d-1}$ for $k=d+2,\dots,n$,
where $\bm{x}=(1,\bm{s}^{\prime})^{\prime}=(1,s_{1},\ldots,s_{d})^{\prime}$ , $\bm{\psi}(\bm{s})=(\psi_{1}(\bm{s}),\dots,\psi_{n}(\bm{s}))'$, and $\bm{v}_{k}$ is the $k$th column of the eigen-vector $\bm{V}$ serving as the coefficients for the linear combination of the $n$ TPS kernels in $\bm{\psi}(\bm{s})$. 
The rationale of the basis construction  procedure is about the smoothness for a function. Let 
\[
J(f)=\int_{R^{d}}\sum_{\gamma_{1}+\cdots+\gamma_{d}=2}\frac{2!}{\gamma_{1}!\cdots\gamma_{d}!}\left(\frac{\partial^{2}f(\bm{s})}{\partial x_{1}^{\gamma_{1}}\cdots\partial x_{d}^{\gamma_{d}}}\right)^{2}d\bm{s}
\] 
measure the roughness of $f$. Consider two function $\tilde{f}$ and  $\check{f}$ satisfying $\sum_{i=1}^{n}\tilde{f}^{2}(\bm{s}_{i})=\sum_{i=1}^{n}\check{f}^{2}(\bm{s}_{i})=1$. 
When $\tilde{f}$ represents a broader scale spatial pattern and $\check{f}$  represents a more local scale spatial pattern, it is expected that $J(\tilde{f})< J(\check{f})$. It has been shown that 
\[
f_{k}={\displaystyle \mathop{\arg\min}_{f}\left\{ J(f):\sum_{i=1}^{n}f(\bm{s}_{i})f_{1}(\bm{s}_{i})=\cdots=\sum_{i=1}^{n}f(\bm{s}_{i})f_{k-1}(\bm{s}_{i})=0,\sum_{i=1}^{n}f^{2}(\bm{s}_{i})=1\right\} };
\]
that is, $f_k(\cdot)$ is the least fluctuating function under the orthogonality constraints and the unit-length identifiability condition. Hence, spatial patterns are extracted sequentially from the smoothest to the roughest in the orthogonal subspaces.  Broader-scale spatial patterns can also be easily captured by $f_k$, in contrast to most TAs that rely only on a few neighbors via neighboring gray-level dependence or gray-level co-occurrence.        

\begin{figure}[t]
\begin{centering}
\includegraphics[scale=0.45]{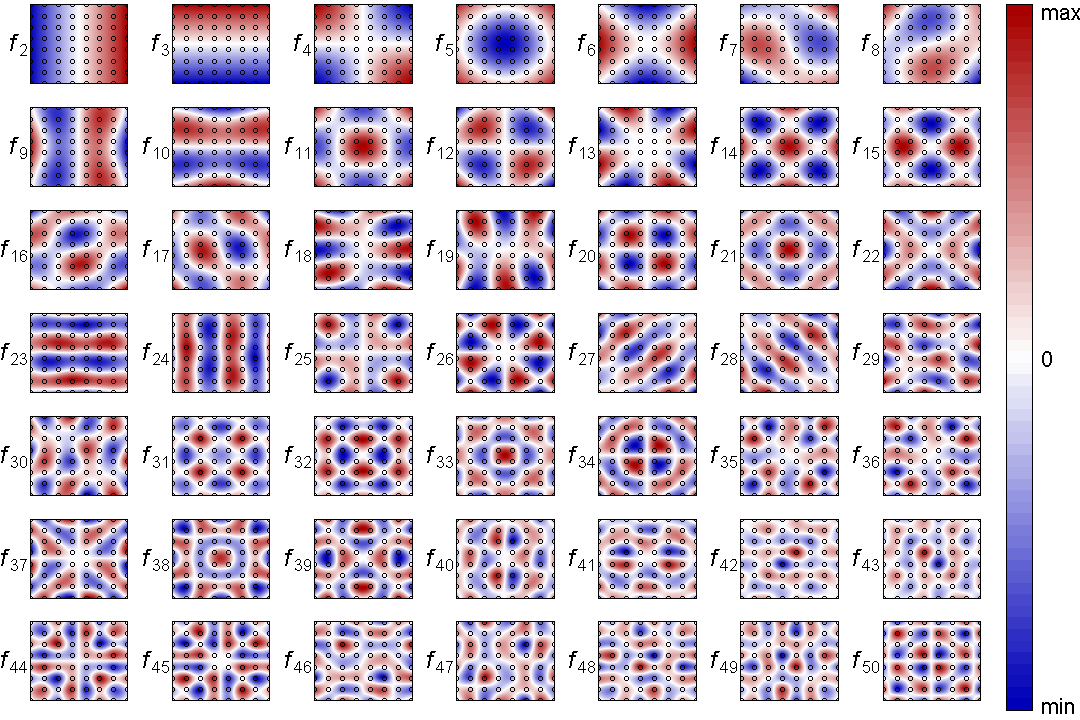}
\par\end{centering}
\caption{An example the basis functions evaluated over $99\times 99$ pixels, superimposed by an $8\times 8$ grid of observation locations (in black circles).}
\label{fig:basis2D}
\end{figure}

An example of the basis functions on an $8\times 8$ image of 64 pixels is given in Figure~\ref{fig:basis2D}.
The multi-resolution basis functions $f_k(\cdot)$ are ordered from the smoothest to the roughest, corresponding to increasing $k$. 
Unlike most multi-resolution basis functions based on wavelets, $f_k(\cdot)$ can be directly applied to irregular grids of observation locations and non-rectangular spatial domains.
It has been shown that using the first $K\leq n$ basis functions, $\{f_{k}(\bm{s});\: k=1,\dots,K\}$, can approximate quite well complex, stationary or nonstationary, covariance functions (\citealp{TzengHuang2017}). 

\subsection{Model Estimation and Selection}\label{sub:estimation and selection}

We estimate the patient-specific mean function by a grand mean 
$$
\hat{\mu}_{j}(\bm{s})\equiv\hat{\mu}_{j}=n_{j}^{-1}\bm{1}_{n_{j}}^{\prime}\mathbf{O}_{j}\bm{z}_{j},
$$
where $\bm{1}_{n_{j}}$ is a vector of $n_j$ 1's.
We let $\tilde{\bm{z}}_{j}=\mathbf{O}_{j}\bm{z}_{j}-\hat{\mu}_{j}$ denote a vector of detrended observations for the $j$th patient; $j=1,\ldots,N$. 

Given the set of basis functions $\{f_{k}(\bm{s});\: k=1,\dots,K\}$ in Section~\ref{sec:basis function}, we develop an expectation-maximization (EM) algorithm for obtaining the maximized likelihood (ML) estimator of $\sigma^{2}$ and $\bm{M}$ and denote them by $\hat{\sigma}^{2}$ and $\hat{\bm{M}}$, respectively \citep{dempster1977EM, katzfuss2011spatio[clm]}.
The random vector  $\bm{w}_{j}$ is treated as unobserved data, and the distribution of $\bm{w}_{j}$ conditional on $\tilde{\bm{z}}_{j}$ is derived in the E-steps, while the estimates of $\sigma^{2}$ and $\bm{M}$ are updated in the M-steps. 

Specifically, for iteration $t=0,1,2,\ldots$, given the estimated parameters $\sigma^{2(t)}$ and $\bm{M}^{(t)}$ in the $t$th iteration, the E-step computes the conditional mean and conditional variance of $\bm{w}_{j}$ as
\begin{align*}
\hat{\bm{w}}_{j}^{(t)} = & E(\bm{w}_{j}|\tilde{\bm{z}}_{1},\ldots,\tilde{\bm{z}}_{n})=\bm{M}^{(t)}\bm{F}_{j}^{\prime}\left\{\bm{F}_{j}\bm{M}^{(t)}\bm{F}_{j}^{\prime}+\sigma^{2(t)}\bm{I}_{n_{j}}\right\}^{-1}\tilde{\bm{z}}_{j},\\
\bm{Q}_{j}^{(t)} = & Var(\bm{w}_{j}|\tilde{\bm{z}}_{1},\ldots,\tilde{\bm{z}}_{n})=\bm{M}^{(t)}-\bm{M}^{(t)}\bm{F}_{j}^{\prime}\left\{\bm{F}_{j}\bm{M}^{(t)}\bm{F}_{j}^{\prime}+\sigma^{2(t)}\bm{I}_{n_{j}}\right\}^{-1}\bm{F}_{j}\bm{M}^{(t)},
\end{align*}
where $\bm{F}_{j}=\bm{O}_j\bm{F}$ and the $(i,k)$th element of the $n\times K$ matrix of basis functions, $\bm{F}$, is $f_{k}(\bm{s}_{i})$ for $i=1,\ldots,n$ and $k=1,\ldots,K$.
Then the M-step updates the parameters as:
\begin{align*}
\bm{M}^{(t+1)} & =N^{-1}\sum_{j=1}^{N}\left\{\hat{\bm{w}}_{j}^{(t)}\left(\hat{\bm{w}}_{j}^{(t)}\right)^{\prime}+\bm{Q}_{j}^{(t)}\right\},\\
\sigma^{2(t+1)} & =\left(\sum_{j=1}^{N}n_{j}\right)^{-1}\sum_{j=1}^{N}\left\{ \tilde{\bm{z}}_{j}^{\prime}\tilde{\bm{z}}_{j}-2\tilde{\bm{z}}_{j}^{\prime}\bm{F}_{j}\hat{\bm{w}}_{j}^{(t)}+\textrm{tr}\left(\bm{F}_{j}\bm{M}^{(t+1)}\bm{F}_{j}^{\prime}\right)\right\} .
\end{align*}
For $t=0$, the estimates are initialized to be $\hat{\bm{w}}_{j}^{(0)}=\bm{0}$ and $\bm{Q}_{j}^{(0)}=\boldsymbol{I}_{K}$.
We alternate between the E- and M-steps iteratively until convergence. 

Let $\hat{\sigma}_{K}^{2}$ and $\hat{\bm{M}}_{K}$ denote the ML estimators of $\sigma^2$ and $\bm{M}$, respectively, with $K$ basis functions.
Also let $\ell(\hat{\sigma}_{K}^{2},\hat{\bm{M}}_{K})$ denote the log-likelihood function for the data model \eqref{eq:data-model} evaluated at $\hat{\sigma}_{K}^{2}$ and $\hat{\bm{M}}_{K}$.
We select $K\in\{d+1,\dots,\tilde{K}\}$ for a sufficiently large $\tilde{K}$ based on the Akaike's information
criterion (AIC), 
$$
\textrm{AIC}(K)=-2\ell(\hat{\sigma}_{K}^{2},\hat{\bm{M}}_{K})+2\mathrm{df}(K).
$$
The degrees of freedom (df) in AIC($K$) is $\mathrm{df}(K)=K(K+1)/2+1$ if $K\leq N$ and $\mathrm{df}(K)=KN+1-N(N-1)/2$ if $K\geq N$; that is, the number of free parameters in $\bm{M}$ depends on the number of basis functions $K$ and the number of patients $N$. 
The number of basis functions selected by AIC is denoted as $K^{*}={\displaystyle \mathop{\arg\min}_{d+1\leq K\leq\tilde{K}}\mathrm{AIC}(K)}$.
An advantage of the selection via AIC is that the computation is more efficient, in contrast to cross validation based methods that are computationally intensive and involve resampling variability \citep{TzengHuang2017}. 

\subsection{Feature Extraction}\label{sec:method:feature}

Although the spatial process $\eta_j(\cdot)$ is composed of component functions as in \eqref{eq:uni-latent-eta}, there is no explicit sample covariance matrix to decompose due to the unequal number of observation locations across patients and the large number of missing values within patients. 
Here, we let the ML estimators of $\bm{M}$ and $\sigma^{2}$ for the final model \eqref{eq:data-model} with $K=K^{*}$ be denoted as $\hat{\bm{M}}_{K^{*}}$ and $\hat{\sigma}_{K^{*}}^{2}$, respectively. 
By $\bm{w}_{j}=[\boldsymbol{u}_{1},\ldots,\boldsymbol{u}_{H}]\bm{\theta}_{j}$, we obtain the estimated component functions $\hat{g}_{h}(\bm{s})$, the estimated number of component functions $H^*$, and the estimated eigen-values $\hat{\lambda}_{h}$, for $h=1,\ldots,H^{*}$, as follows.  

Let $\hat{\boldsymbol{U}}\mathrm{diag}\{\hat{\lambda}_{1},\dots,\hat{\lambda}_{K^{*}}\}\hat{\boldsymbol{U}}'$
be the eigen-decomposition of $\hat{\bm{M}}_{K^{*}}$, where $\hat{\boldsymbol{U}}=[\hat{\boldsymbol{u}}_{1},\ldots,\hat{\boldsymbol{u}}_{K^{*}}]$
and $\hat{\lambda}_{1}\geq\cdots\geq\hat{\lambda}_{H^*}>0=\hat{\lambda}_{H^*+1}=\cdots=\hat{\lambda}_{K^{*}}$.
We only need the estimates of the first $H^*$ components corresponding to the positive eigen-values.
Then by (\ref{eq:sum-of-basis-functions}), we estimate $g_h(\bm{s})$, for $h=1,\ldots,H$, by 
\begin{equation}
\hat{g}_{h}(\bm{s})=\hat{\bm{u}}_{h}^{\prime}\boldsymbol{f}_{K^{*}}(\bm{s}).
\label{eq:eigen-function-formula}
\end{equation}
In addition, an estimator of the covariance between two locations for the $j$th patient is 
$$
\widehat{\textrm{cov}}(y_{j}(\bm{s}),y_{j}(\bm{s}^*))=\sum_{h=1}^{H^*}\hat{\lambda}_{h}\hat{g}_{h}(\bm{s})\hat{g}_{h}(\bm{s}^*),
$$ 
where $\hat{g}_{h}(\bm{s})$ is the $h$th estimated component function \eqref{eq:eigen-function-formula}.

Further, let $\hat{\bm{\Lambda}}=\mathrm{diag}\{\hat{\lambda}_{1},\dots,\hat{\lambda}_{H^*}\}$ denote the estimated variance matrix for $\bm{\theta}_j$.
We define an $n_{j}\times H^*$ matrix $\hat{\bm{G}_{j}}$ for the $H^*$ component functions taken values at observation locations for the $j$th patient, with the $(i,h)$th element being $\hat{g}_{h}(\bm{s}_{j,i})$ for $i=1,\ldots,n_j, h=1,\ldots,H^*$.
Thus, an empirical predictor with the smallest mean squared prediction error (MSPE) of $y_{j}(\bm{s})$ is given by 
\begin{align}
\hat{y}_{j}(\bm{s}) & =\hat{\mu}_{j}+\mathrm{\widehat{cov}}\left(y_{j}(\bm{s}),\mathbf{O}_{j}\bm{z}_{j}\right)\mathrm{\widehat{cov}^{-1}}\left(\mathbf{O}_{j}\bm{z}_{j},\mathbf{O}_{j}\bm{z}_{j}\right)\left(\mathbf{O}_{j}\bm{z}_{j}-\hat{\mu}_{j}\right)\nonumber \\
 & =\hat{\mu}_{j}+\hat{\bm{g}}(\bm{s}){}^{\prime}\hat{\boldsymbol{\Lambda}}\hat{\bm{G}_{j}}^{\prime}\left(\hat{\bm{G}}_{j}\hat{\boldsymbol{\Lambda}}\hat{\bm{G}_{j}}^{\prime}+\hat{\sigma}_{K^{*}}^{2}\bm{I}_{n_{j}}\right)^{-1}\tilde{\bm{z}}_{j},
 \label{eq:yhat}
\end{align}
where  $\hat{\bm{g}}(\bm{s})=(\hat{g}_{1}(\bm{s}),\ldots,\hat{g}_{H^*}(\bm{s}))^{\prime}$.
It follows that the conditional expectation of $\bm{\theta}_{j}$ given data can be obtained by equating $\hat{y}_{j}(\bm{s})=\hat{\mu}_{j}+\sum_{h=1}^{H}\hat{\theta}_{jh}\hat{g}_{h}(\bm{s})=\hat{\mu}_{j}+\hat{\bm{g}}(\bm{s})^{\prime}\hat{\bm{\theta}}_{j}$ with (\ref{eq:yhat}) and thus,
\begin{equation}
\hat{\bm{\theta}}_{j}=\hat{\boldsymbol{\Lambda}}\hat{\bm{G}_{j}}^{\prime}\left(\hat{\bm{G}}_{j}\hat{\boldsymbol{\Lambda}}\hat{\bm{G}_{j}}^{\prime}+\hat{\sigma}_{K^{*}}^{2}\bm{I}_{n_{j}}\right)^{-1}\tilde{\bm{z}}_{j}.
\label{eq:feature-formula}
\end{equation}
We then consider the vector $\left(\hat{\mu}_{j},\hat{\bm{\theta}}_{j}^{\prime}\right)^{\prime}$ to be the data-driven features for the $j$th patient. 

For a new patient with image intensity data denoted as $\bm{z}_0$ observed at $n_0$ locations $\bm{r}_1,\ldots,\bm{r}_{n_0}$, the data-driven features can be computed as follows. 
First, the patient-specific mean function is estimated to be $\hat{\mu}_0=n_0^{-1}\bm{1}_{n_0}^{\prime}\bm{z}_0$ and is subtracted from $\bm{z}_0$ to obtain $\tilde{\bm{z}}_0$. 
We can then construct $\hat{\bm{G}}_0$, whose $(i,h)$th element is $\hat{g}_{h}(\bm{r}_i)$ in (\ref{eq:eigen-function-formula}) for $i=1,\ldots,n_0, h=1,\ldots,H^*$. 
Together with the estimated $\hat{\boldsymbol{\Lambda}}$ and $\hat{\sigma}_{K^{*}}^{2}$, the estimated random weights $\hat{\bm{\theta}}_0$ are obtained from (\ref{eq:feature-formula}) with $j=0$. 

%
%
\section{Simulation Study}\label{sec:simu}

\begin{figure}[h]
\begin{centering}
\includegraphics[scale=0.55]{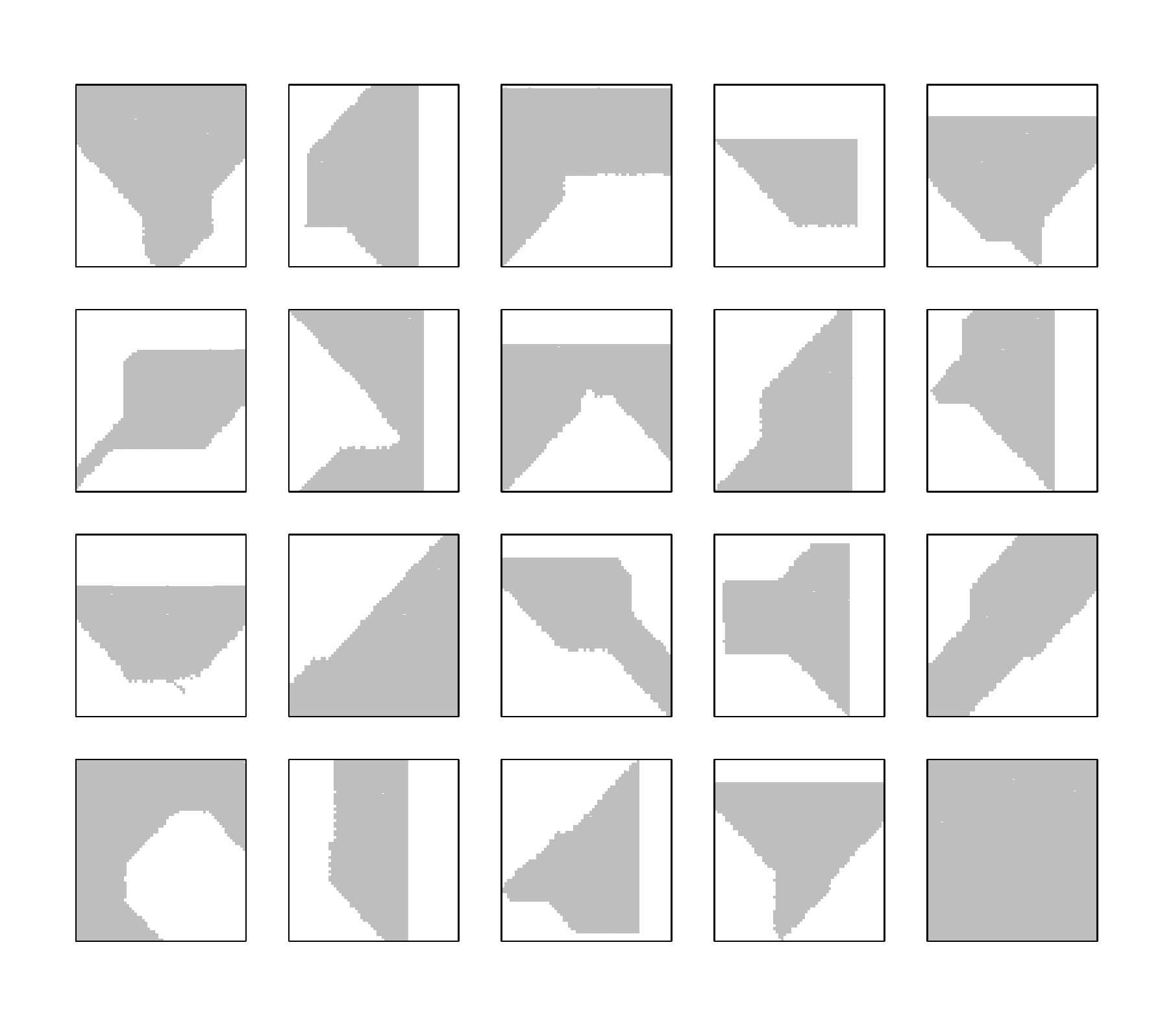}
\par\end{centering}
\caption{Simulated ROIs (gray area) in a unit square image across $N=20$ patients, where the white space within each image is viewed as missing.}
\label{fig:2Dsamples}
\end{figure}

We conducted a simulation study to investigate the properties of the proposed methodology in Section~\ref{sec:method}, focusing on the determination of component functions.
Four sample sizes (i.e., number of patients) $N=30, 40, 50$, and $60$ were considered and the image for any given patient was in the unit square ${\cal U}=[0,1]\times[0,1]$.
The number of pixels (or, image resolution) in the unit square was set to $n=2^{L}\times 2^L$, where $L=3, 4, \ldots, 7$, corresponding to the total number of pixels $n=2^3\times 2^3, 2^4\times 2^4, \ldots, 2^7\times 2^7$, respectively. 
Given a sample size $N$ and an image resolution, different shapes of ROIs were generated within the square for $N$ different patients, where the proportion of pixels outside the ROI (i.e., the missing rate) was selected to be between 0.25 and 0.6 mimicking the data example in Section~\ref{sec:example}. 
Figure~\ref{fig:2Dsamples} illustrates the simulated ROIs for $N=20$ patients and $n=64 \times 64 $ pixels per image. 

The true image intensity, or signal, for the $j$th patient was generated according to a linear combination of cosine functions:
\[
y_{j}(\bm{s})=\sum_{k=1}^{17}\delta_{k}w_{jk}\cos(\left\Vert \bm{s}-\bm{c}_{k}\right\Vert ),
\]
where the cosine functions $\cos(\left\Vert \bm{s}-\bm{c}_{k}\right\Vert )$ for $\bm{s}\in{\cal U}$ relied on the Euclidean distance $\left\Vert \bm{s}-\bm{c}_{k}\right\Vert$ between $\bm{s}$ and 17 pre-selected center points defined to be $\bm{c}_{k}\in\{(i/5,\: i^{*}/5)^{\prime};\: i,i^{*}=1,2,3,4\}\cup\left\{ (0.5,0.5)^{\prime}\right\} $.
The random weights for the cosine functions were independently distributed as $w_{jk}\sim\mathcal{N}\left(0,(\tau+\varsigma_{k})\right)$, where $\tau$ controlled the signal-to-noise ratio (SNR) and  $\varsigma_{k}$ introduced heterogeneity for the variances of different components.
We selected $\tau$ to be 1, 2, 3, or 4 and $\varsigma_{k}$ was drawn independently and uniformly from the interval $\left[-0.5,0.5\right]$.
Further, $\delta_{k}=1$ if the $k$th cosine function was included in the model and $\delta_{k}=0$ otherwise. 
That is, only a portion of the cosine functions were used in generating $y_{j}(\bm{s})$, determined by an image complexity level $p=\sum_{k=1}^{17}\delta_{k}$.
We considered seven image complexity levels, $p=0,1,\ldots,6$. 
An observation of intensity $z_{j}(\bm{s})$ in the pixel centered at $\bm{s}$ was then generated according to $z_{j}(\bm{s})=y_{j}(\bm{s})+\varepsilon_{j}(\bm{s})$, where the errors followed the standard normal distribution $\varepsilon_{j}(\bm{s})\stackrel{iid}{\sim}\mathcal{N}(0,1)$. 

\begin{figure}
\noindent \begin{centering}
\includegraphics[scale=0.70]{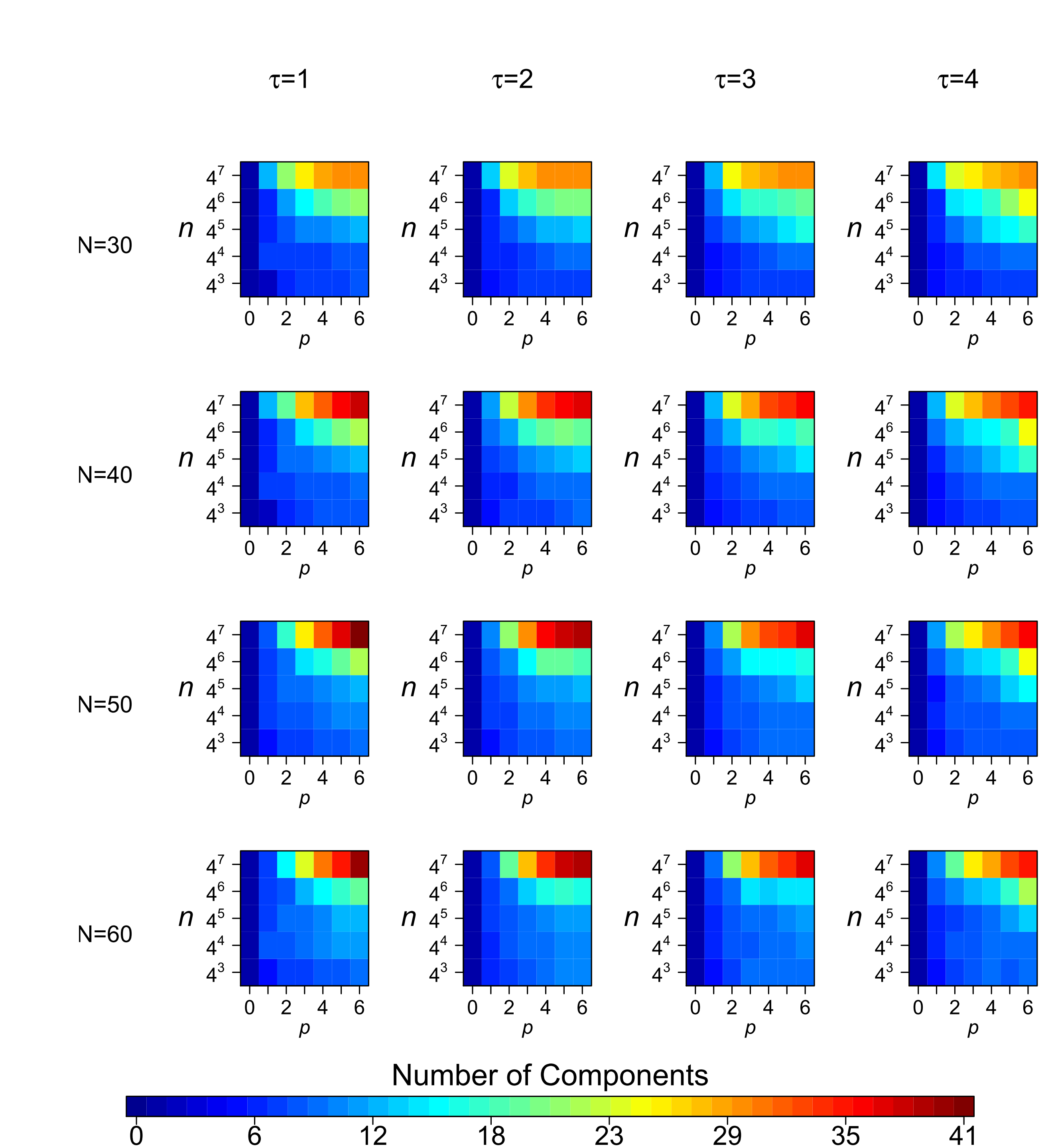}
\par\end{centering}
\caption{Median estimated numbers of extracted component functions for simulated ROIs from $N=30,40,50,60$ patients (across rows), $\tau=1,2,3,4$ signal-to-noise ratios (across columns), $n=2^3\times 2^3, \ldots, 2^7\times 2^7$ image resolutions ($y$-axis within an image), and $p=0,1,\ldots,6$ image complexity levels ($x$-axis within an image).}
\label{fig:4x4}
\end{figure}

In total, there were 560 combinations of the number of patients $N$, the image resolution $n$, the SNR factor $\tau$, and the image complexity level $p$. 
For each combination, we ran 100 simulations and computed the median of the estimated numbers of extracted component functions, $H^*$, given in Section~\ref{sec:method:feature}. 
Since the cosine functions are not orthogonal to each other, the number of component functions was not necessarily the same as the image complexity level $p$. 
Moreover, by \eqref{eq:sum-of-eigen-functions}, it could happen that an underlying cosine function be approximated by more than one estimated component functions. As shown in Figure~\ref{fig:4x4}, the patterns of $H^*$ for different sample sizes $N$ and SNR factors $\tau$ are quite similar. 
The determining factors seem to be $n$ and $p$ such that as the image resolution increases and/or the image complexity increases, a larger number of component functions would be extracted.
These results are desirable, as in practice, the true eigen-functions are unknown and we expect to be able to extract more information when the underlying image structure is more complex or when the image resolution is higher with more and finer details.  

%
%
\section{Data Example}\label{sec:example}

\subsection{Cancer Image Data}\label{sec:example:data}

In a cancer research study, \citet{bradshaw2015molecular} collected 3D image data from canine patients with spontaneous sinonasal tumors, investigating the association between clinical end points of interest and potential imaging biomarkers.
\citet{bradshaw2015molecular} adopted positron emission tomography (PET) as the imaging technique, which injects a radioactive molecule into an individual patient and records where specific molecular functions are occurring. 
PET images were also paired with computed tomography (CT) images, providing both functional and anatomic information about the patient. 
Although pre-, mid-, and post-treatment data were available, here we focus on the pre-treatment data to extract QIBs for prognostic purposes.

Three PET radiotracers were considered for each patient in this data example, namely, fluorodeoxyglucose (FDG) which measures glucose metabolism, flurothymidine (FLT) which measures proliferating cells, and Cu(II)-diacetyl-bis(N4-methylthiosemicarbazone) (Cu-ATSM) which measures hypoxia. 
All measurements were normalized by the dose injected into the patient and the body weight of the patient to achieve standardized uptake values (SUVs) for subsequent analysis. For each patient, the ROI is the contoured gross tumor volume.
The SUVs outside the ROI were regarded as missing, as it was reasonable to assume that the uptakes inside and outside the ROI were very different and only those uptakes inside the ROI were the target of interest. 
Figure~\ref{fig:zandg}(a) illustrates an example of the cancer ROI for one of the canine patients.

\subsection{Feature Extraction and Identification}

We applied the methodology developed in Sections~\ref{sec:model}-\ref{sec:method} to extract QIBs based on the 3D cancer image data described in Section~\ref{sec:example:data}.
In particular, we had three images for each of the $N=22$ canine patients for the three PET radiotracers, FDG, FLT, and Cu-ATSM. 
The $j$th patient had an ROI with $n_{j}\leq n$ voxels across the three radiotracers and the images had missing values outside the ROIs. 
For each of the $N=22$ patients, the centroid of each ROI was transformed linearly to be the origin $(0,0,0)^{\prime}$ in $\mathbb{R}^{3}$. 
The union of the transformed voxel centroids for all the ROIs formed the set $\left\{ \bm{s}_{1},\dots,\bm{s}_{n}\right\} $ where the total number of voxels in $D$ was $n=19697$. 

Model fitting was performed for each radiotracer separately based on the model given in Section~\ref{sec:model}, where the image data were the SUVs at transformed voxel locations within each patient's ROI. 
The model parameters were estimated by the EM algorithm and the number of basis functions $K$ was selected by the AIC as described in Section~\ref{sub:estimation and selection}. 
Then, the number of component functions $H$ and the component functions $g_h(\cdot)$ were estimated following Section~\ref{sec:method:feature}. 
We had $K^*_{\rm FDG}=855$, $H^*_{\rm FDG}=18$ for FDG, $K^*_{\rm FLT}=540$, $H^*_{\rm FLT}=13$ for FLT, and $K^*_{\rm Cu-ATSM}=595$, $H^*_{\rm Cu-ATSM}=16$ for Cu-ATSM. 
Figure~\ref{fig:zandg}(b) illustrates one of the estimated component functions $\hat{g}_h(\cdot)$.

\begin{figure}
\begin{centering}
\includegraphics[scale=0.65]{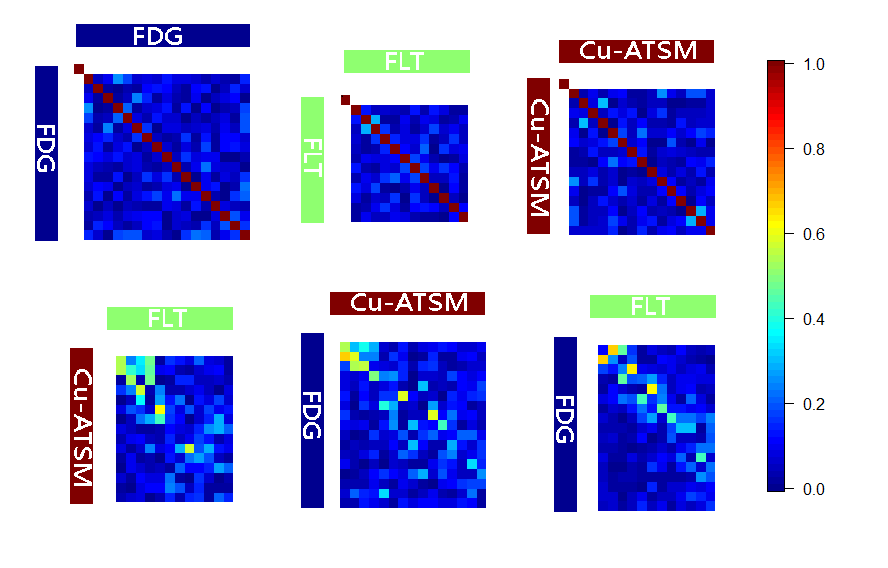}
\par\end{centering}
\caption{Absolute correlations between the estimated component functions $\hat{g}_{h}$ and $\hat{g}_{h^*}$ over the observation locations $\bm{s}_{1},\dots,\bm{s}_{n}$ for radiotracers FDG and FDG (top left), FLT and FLT (top middle), Cu-ATSM and Cu-ATSM (top right), FLT and Cu-ATSM (bottom left), FDG and Cu-ATSM (bottom middle), FLT and FDG (bottom right).}
\label{fig:corr-matrix}
\end{figure}

\begin{figure}
\begin{centering}
\includegraphics[width=10.8cm,height=6.4cm]{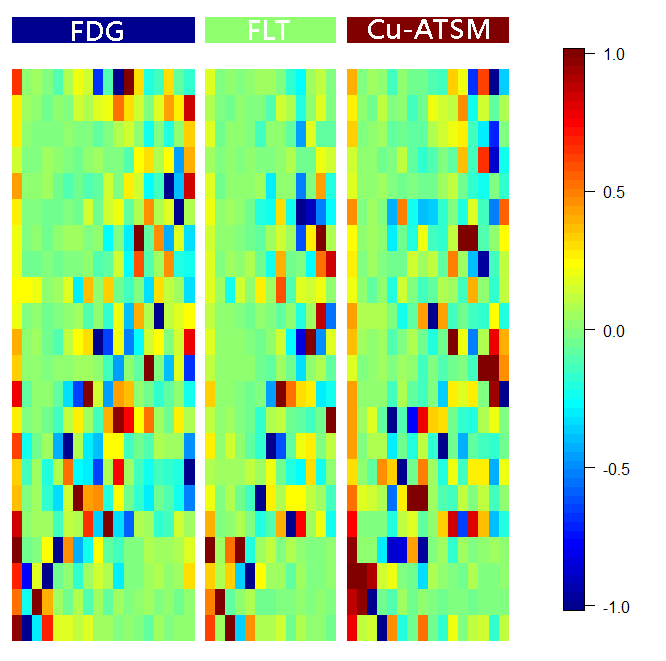}
\par\end{centering}
\caption{Estimated data-driven features $\hat{\bm{{\theta}}}_{j}^{\rm FDG}$, $\hat{\bm{{\theta}}}_{j}^{\rm FLT}$,
and $\hat{\bm{{\theta}}}_{j}^{\rm Cu-ATSM}$ for patients $j=1,\ldots,22$ (rows) and for the three radiotracers, FDG (left), FLT (middle), and Cu-ATSM (right). For each radiotracer, the columns were sorted by eigenvalues and the values plotted in the heat maps were scaled by the largest absolute correlation for visual comparison.}
\label{fig:pc-score}
\end{figure}

The pairwise (Pearson) correlations between the estimated component functions $\hat{g}_{h}(\bm{s})$ are shown in Figure~\ref{fig:corr-matrix}. 
Within each radiotracer, the correlation between the estimated component functions was generally small, as was expected due to orthogonality.
Between radiotracers, however, there was substantial correlation with an average of 72.3\% of the component functions for a given radiotracer to have more than 0.5 absolute correlation with the other two radiotracers. 
Although the metabolic mechanisms of the three radiotracers were very different, their underlying image characteristics were quite related as these correlations between different component functions reflected. 
In addition, for the $j$th patient, the estimated data-driven features $\left\{ \hat{\theta}_{jh}^{\rm tracer}: h=1,\ldots,H^*_{\rm tracer}\right\}$ are plotted in the $j$th row, for $j=1,\ldots,22$, in Figure~\ref{fig:pc-score}, where the tracer is FDG, FLT, or Cu-ATSM. 
The patterns of the estimated data-driven features are similar across the three radiotracers.

\subsection{Comparison of Methods}

We compared three sets of QIB features extracted from the 3D cancer image data: (1) features extracted by our proposed method based only on pre-treatment images; (2) the pre-treatment and mid-treatment features used in \citet{bradshaw2015molecular}; and (3) features extracted from a commonly used TA method as described in \citet{galavis2010variability} based only on pre-treatment images. 
\citet{bradshaw2015molecular} found two features, the mid-treatment maximum FLT SUV and the relative change between pre- and mid-treatment mean FLT SUV, to be useful out of 26 candidate features considered in the association with a clinical end point, progression-free survival time, defined as the time from treatment initiation to any one of the events of metastasis, tumor growth, or recurrence. 
The third set by TA had 50 features, including 8 first-order features containing no information about the relative position of the voxels, 28 second-order features (with 23 features using a co-occurrence matrix and 5 features using the neighboring gray level), and 14 higher-order features (with 11 features employing a gray level run length matrix and 3 features using the neighborhood gray tone difference matrix). 
Table~\ref{tab:feature correlation} compares the distributions of absolute correlations between features extracted by our method and those by TA, showing clearly that our method had smaller correlations within and between features.

We also considered the association of three sets of features with progression-free survival time.
The first set based on our method had the data-driven features shown in Figure~\ref{fig:pc-score}.
The second set extracted by \citet{bradshaw2015molecular} had the 26 features therein, while the third set by TA had $50\times3=150$ features from three radiotracers.
\citet{bradshaw2015molecular} only considered linear combinations of covariates in a Cox proportional hazards model.
To include nonlinearity, we applied a tree-based algorithm for survival data using the R package $\mathtt{rpart}$ (\citealp{radespiel2003comparison}). 
By the rule of thumb of needing 10 observations per covariate, we selected the best model based on a pseudo-$\textrm{R}^{2}$ among the Cox proportional hazards models with up to two features. 

The model fitting results are as follows. 
\citet{bradshaw2015molecular}'s method had a pseudo-$\textrm{R}^{2}$ value of 0.53 and the AIC value was 72.21. 
For the purpose of developing prognostic QIBs, we did not use the mid-treatment information for our method and the TA method. 
The TA method achieved a pseudo-$\textrm{R}^{2}$ value of 0.52 and the AIC value was 73.17, which were approximately the same as \citet{bradshaw2015molecular}, although with only pre-treatment information. 
Our proposed method had the highest pseudo-$\textrm{R}^{2}$ value of 0.63 and the smallest AIC of 67.16 among the three methods, indicating an improvement over the existing feature extraction methods.

As mentioned in Section~\ref{sec:intro}, quantization could have a large impact on the features extracted by TA. 
We conducted a sensitivity analysis by recalculating the 150 features by the TA method with numbers of bins different from the default of 256 bins used above.
We examined the correlations between the features based on the default 256 bins and the recalculated features based on 128, 64, 32, and 16 bins, shown in Table~\ref{tab:quantization}.
Although the pseudo-$\textrm{R}^{2}$ and AIC values were similar across the numbers of bins, the recalculated features became less correlated with the features based on 256 bins, as the number of bins decreased.
These results demonstrated the challenges associated with quantization in the conventional TA method, as the results were sensitive to the choice of the bin number and it was unclear how to select an appropriate bin number. 

%
%
\section{Conclusions and Discussion}\label{sec:conclusion}

In this paper, we have developed a general and flexible statistical approach for extracting QIBs from medical images in an objective and principled way. 
New statistical methodology has been developed based on spatial process decomposition where the random weights are unique to individual patients for component functions common across patients. Model fitting, model selection, and feature extractions have been demonstrated using both simulated image data and an actual cancer image data set. 
Medical images have inherent uncertainties that play an important role in the ability to develop useful QIBs. Recent works have shown that the conventional TA techniques are sensitive to image reconstruction techniques, patient motion during scanning, ROI definitions, among others \citep[see Table~1.1,][]{harmon2016molecular}.
The method here is not a solution for all the uncertainties in PET QIB development, but does have advantages over the existing methods including approximate orthogonality among the extracted features, no need for \textit{ad-hoc} quantization of features, and automatic tuning parameter selection. In addition, our method can handle non-rectangular ROIs with varying shapes among patients and applies to 1D curve, 2D pixel, and 3D voxel data in a unified way.

\bibliography{reference}

\section*{Tables and Figures}

\begin{table}[th]
\begin{centering}
\begin{tabular}{lccccc}
\hline 
Absolute & Method & \multicolumn{2}{c}{Within Radiotracer} & \multicolumn{2}{c}{Between Radiotracer}\tabularnewline
Correlation  & \citet{bradshaw2015molecular} 
& Our  & TA  & Our  & TA\tabularnewline
\hline 
$[0, 0.1)$  & 34.1\% & 78.9\%  & 3.2\%  & 70.1\%  & 50.3\%\tabularnewline
$[0.1, 0.3)$  & 21.3\% & 15.3\%  & 8.2\%  & 20.9\%  & 5.3\%\tabularnewline
$[0.3, 0.5)$  & 20.6\% & 3.0\%  & 23.6\%  & 4.8\%  & 10.3\%\tabularnewline
$[0.5, 0.7)$  & 14.0\% & 2.0\%  & 21.0\%  & 1.9\%  & 11.6\%\tabularnewline
$[0.7, 0.9)$  & 8.3\% & 0.0\%  & 7.8\%  & 2.2\%  & 4.6\%\tabularnewline
$[0.9,1.0)$  & 1.7\% & 0.8\%  & 36.2\%  & 0.1\%  & 17.9\%\tabularnewline
\hline 
\end{tabular}
\par\end{centering}
\caption{Percentages of the absolute correlations between features extracted by three methods: our proposed method, \citet{bradshaw2015molecular}, and textural analysis (TA) within and between radiotracers.}
\label{tab:feature correlation} 
\end{table}

\begin{table}
\begin{centering}
\begin{tabular}{lrrrr}\hline 
Absolute & \multicolumn{4}{c}{Number of Bins}\tabularnewline
Correlation & 128 & 64 & 32 & 16 \tabularnewline
\hline 
$[0, 0.5)$ & 0\% & 0\% & 2.7\% & 10.0\%\tabularnewline
$[0.5, 0.9)$ & 0.7\% & 2.0\% & 14.7\% & 26.7\%\tabularnewline
$[0.9, 0.95)$ & 0\% & 4.0\% & 9.3\% & 9.3\%\tabularnewline
$[0.95, 0.99)$ & 8.7\% & 22.7\% & 24.7\% & 12.7\%\tabularnewline
$[0.99, 1.0)$ & 90.7\% & 71.3\% & 48.7\% & 41.3\%\tabularnewline
\hline 
pseudo-$\textrm{R}^{2}$ & 0.54 & 0.55 & 0.50 & 0.50\tabularnewline
AIC & 72.23 & 71.61 & 74.09 & 73.77\tabularnewline
\hline 
\end{tabular}
\par \end{centering}
\caption{Percentages of the absolute correlations between features extracted by TA with 256 bins and those with 128, 64, 32, and 16 bins, and the corresponding pseudo-R$^2$ and AIC values.}
\label{tab:quantization}
\end{table}

\end{document}